\documentclass[11pt]{article}

\usepackage{common}

\title{Minimum Average Deviance Estimation for Sufficient Dimension Reduction}
\author{Kofi~P. Adragni$^{a*}$, Andrew~M. Raim$^b$, \& Elias Al-Najjar$^a$ \vspace{0.5em} \\
$^a$Department of Mathematics and Statistics, University of Maryland, Baltimore County \\
$^b$Center for Statistical Research and Methodology, U.S. Census Bureau}
\date{}

\begin{document}
\maketitle
\blfootnote{$^*$Corresponding author: \url{kofi@umbc.edu}}
\blfootnote{Disclaimer: This article is released to inform interested parties of ongoing research and to encourage discussion of work in progress. Any views expressed are those of the authors and not necessarily those of the U.S.~Census Bureau.}

\begin{abstract}
Sufficient dimension reduction reduces the dimensionality of data while preserving relevant regression information. In this article, we develop Minimum Average Deviance Estimation (MADE) methodology for sufficient dimension reduction. It extends the Minimum Average Variance Estimation (MAVE) approach of Xia et al. (2002) from continuous responses to exponential family distributions to include Binomial and Poisson responses. Local likelihood regression is used to learn the form of the regression function from the data. The main parameter of interest is a dimension reduction subspace which projects the covariates to a lower dimension while preserving their relationship with the outcome. To estimate this parameter within its natural space, we consider an iterative algorithm where one step utilizes a Stiefel manifold optimizer. We empirically evaluate the performance of three prediction methods, two that are intrinsic to local likelihood estimation and one that is based on the Nadaraya-Watson estimator. Initial results show that, as expected, MADE can outperform MAVE when there is a departure from the assumption of additive errors.
\end{abstract}

\section{Introduction}

Consider the classical regression problem of estimating the mean function $E(Y \mid X)$ with $Y \in \Rbb$ and $X \in \Rbb^{p}$. When $p$ is large, it is often worthwhile to reduce the dimensionality of $X$ for a better graphical exploration of the data, more parsimonious modeling, and more efficient prediction. The need for dimension reduction methods has increased in past decades to deal with high dimensional data that arise frequently in modern scientific research.

A number of approaches for dimension reduction exist in the literature. Of interest in regression is the concept of sufficient dimension reduction defined by \cite{Cook07} as follows. A reduction $R: \Rbb^p \rightarrow \Rbb^d, d \leq p,$ is sufficient if it satisfies one of the following three statements: (i.) $Y \mid X \sim Y \mid R(X)$, (ii.) $X \mid (Y, R(X)) \sim X \mid R(X)$, or (iii.) $X \indep Y \mid R(X)$. The symbol $\indep$ denotes statistical independence, and $U \sim V$ denotes $U$ and $V$ having identical distributions. Typically, the sufficient reduction $R(X)$ has a dimension strictly less than $p$, so that a more parsimonious analysis is possible by replacing the predictor $X$ with $R(X)$ in the regression of $Y$ on $X$.

Many methods for sufficient dimension reduction exist. They can be roughly grouped into three classes. Moment-based methods include sliced inverse regression \citep[SIR;][]{Li91}, inverse regression estimation \citep[IRE;][]{CookN05}, contour regression \citep{LiZC05}, and directional regression \citep[DR;][]{LiW07}. Likelihood-based methods include principal fitted components \citep[PFC;][]{Cook07} and likelihood acquired directions \citep[LAD;][]{CookF09}. Kernel-based methods include MAVE \citep{Xia02}, Penalized MAVE \citep{WangXiZhu2013} and other variants.

The majority of sufficient dimension reduction methods assume that $X$ is random and $Y$ is fixed; these are known as inverse regression methods. Nearly all moment-based methods are based on the first few moments of $X \mid Y$ and likelihood-based methods assume a distribution of $X \mid Y$. Arguably, inverse regression methods are best suited to deal with high dimensionality of the predictors. However, forward regression methods such as MAVE \citep{Xia02}, which assume $Y$ is random and $X$ is fixed, have been developed with a great deal of success. MAVE is based on the estimation of the gradient of the conditional expectation $E(Y \mid X)$ by way of a local regression. It does not impose strong assumptions on the distribution of $X$, and is developed essentially for continuous $Y$ to yield the so-called central mean subspace \citep{Cook98}.

In this paper, our focus is on regressions with exponential family response $Y$, including Binomial, Poisson, Geometric, Negative Binomial, Gaussian, Exponential, Gamma, Inverse Gamma, and Log-normal distributions. Our goal is to obtain a sufficient dimension reduction $R(X)$ so that $Y \mid X \sim Y \mid R(X)$ without much assumption on $X$. We proceed using a local regression method. Local regression methods have been well-established in the literature. A well-known example is the local likelihood estimation of \citet{TibshiraniHastie1987}. \citet{FanGibjels1996} and \citet{Loader99} provide introductions to local regression and local likelihood. These methods carry out inference at a given $X_0$ using a locally weighted log-likelihood. The contribution of observed $(Y,X)$ may be weighted through a kernel density function whose bandwidth controls the degree of localization. Local likelihood methods allow the fit to vary locally at each $X_0$ of interest, and are also capable of estimating the relationship between $Y$ and $X$ without full specification of a parametric form for the regression function.

\cite{Lambert06} developed a local likelihood regression in generalized linear models. Their work focuses essentially on a single-index reduction. Ours expands to multiple-index to obtain the sufficient dimension reduction of dimension $d$ which is to be estimated. In MAVE, the reduction matrix of dimension $d$ in $\Rbb^p$ was estimated one column at the time, while fixing the other columns in iterations between two quadratic optimizations steps. We instead consider optimization over the natural parameter space of the reduction matrix, which is either a Stiefel or Grassmann manifold, depending whether remaining parameters are held fixed. We also explore three prediction methods to be used with the sufficient reduction without requiring a parametric model.

The remainder of the article is structured as follows. Section~\ref{sec:llad} presents the MADE methodology, including the model and estimation procedure. Section~\ref{sec:dim} discusses inference methods for the dimension $d$ of the reduction subspace. Section~\ref{sec:sim} we presents the three prediction methods and provides a simulation study comparing their performance. Simulations are also carried out to demonstrate effectiveness to estimate the reduction. Some applications to datasets are given in section~\ref{sec:appl} and section~\ref{sec:discuss} ends with concluding remarks.

\section{Minimum Average Deviance Estimation}
\label{sec:llad}

\subsection{Model}
Suppose $Y \in \Rbb$ is a response, $X$ is a $p$-dimensional predictor, and the distribution of $Y \mid X$ is given by an exponential family distribution of the form
\begin{equation}
f(Y \mid \vartheta (X) ) = f_0(Y, \phi) \exp\left\{ [Y \vartheta(X) - b(\vartheta(X))]/a(\phi) \right\}.
\label{eqn:exp-fam}
\end{equation}
For a particular dataset, a specific form of distribution \eqref{eqn:exp-fam} would be assumed, yielding the functions $a(\cdot), b(\cdot),$ and $f_0(\cdot. \cdot)$. The canonical parameter $\vartheta(X)$ is related to the mean function $E(Y \mid X)$ through a link function $g$ so that $g(E(Y \mid X))= \vartheta(X)$. The variance function $\var(Y \mid X) = a(\phi) b^{''}(\vartheta(X))$ where $\phi$ is referred to as dispersion parameter. It is often assumed that $\vartheta(X) = \alpha + \beta^T X$ which amounts to the generalized linear model \citep{McCullaghNelder1989}. The specific distribution of $Y \mid X$ determines the choice of the link function $g(\cdot)$ which is assumed to be the canonical link for the remainder of the paper.

The canonical parameter $\vartheta(X)$ holds the main information that connects $Y$ to $X$. Let $(Y_i, X_i), i=1, \ldots, n$ represent independent samples from the distribution of $(Y, X)$ so that $Y_i \mid X_i$ has the distribution \eqref{eqn:exp-fam}. We will assume that $\vartheta(X)$ is a continuous and smooth function so that it admits at any point $X$ the first order linear expansion
\begin{equation}
\vartheta(X_i) \approx \vartheta(X) + [\nabla\vartheta(X)]^T(X_i-X)
\label{eqn:theta-approx}
\end{equation}
for any $X_i$ in the neighborhood of $X$. Let $\alpha = \vartheta(X)$ and $\Gamma = \nabla\vartheta(X)$. The term $\Gamma$ retains the core information that connects $Y_i$ to $X_i$ locally at $X$. As $X$ varies in its sample space, $\Gamma$ describes a $u$-dimensional subspace $\Scal$ in $\Rbb^p$ with  $u \leq \min(n, p)$. Let $U$ be an orthonormal basis of $\Scal$ so that $\Gamma = U \delta$ for some $\delta = \delta(X) \in \Rbb^{u \times 1}$. It follows that $\vartheta(X_i)=\vartheta(U^TX_i)$, thus the distribution of $Y \mid X$ is approximately the same as that of $Y \mid U^TX$. Consequently, $U^TX$ can be used in lieu of $X$ in the regression of $Y$ on $X$.

The subspace $\Scal$, called a sufficient dimension reduction subspace \citep{Cook07}, is not unique, and may not be minimal in its dimension. When the dimension of $\Scal$ is $p$, no reduction is achieved. Of all the sufficient dimension reduction subspaces, let $\Scal_B$ be the subspace of minimal $d$ with $0 < d \leq u \leq \min(n,p)$ with basis matrix $B$ (which is one of many possible bases). Then $B$ is a semi-orthogonal matrix whose columns span $\Scal_B$. Consequently, we have $\vartheta(X) = \vartheta(B^TX)$ and $X \in \Rbb^p$ can be replaced by $B^T X \in \Rbb^d$ in the regression of $Y$ on $X$. For single-parameter distributions such a Binomial or a Poisson, $\Scal_B$ is a central subspace. However, for a two-parameter distribution such as Gaussian, $\Scal_B$ is a central mean subspace.

We will proceed under the assumption of a one-parameter exponential family so that the dispersion parameter $\phi$ is known. This will be sufficient to develop MADE under Gaussian, Poisson, and Binomial families, which have been the main focus of this work. We will also discuss in section~\ref{sec:twopar} on extensions needed to estimate an unknown $\phi$, to facilitate the use of other useful family types.

The ultimate goal is to determine the reduction subspace $\Scal_B$. A typical way is to determine $B$ so that locally at each point $X_i$, $Y_i$ is the closest to $E(Y_i \mid X_i)=g^{-1}(\vartheta(B^TX_i))$ for all $(Y_i, X_i)$. Consider for example $Y$ from a normal distribution where the link function $g(.)$ is the identity function. The closeness of $Y$ to $E(Y_i \mid X_i)$ can be evaluated with a square loss function, and consequently, the parameter $B$ can be estimated as
$$\Bhat = \argmin_{B \in St(d,p)} E\{E[Y - E(Y \mid B^T X)]^2|B^TX\}$$
where $St(d,p)$ is a Stiefel manifold, the set of all $d$-dimensional orthonormal matrices in $\Rbb^{p}$. For a discrete $Y$ from a Bernoulli distribution for example, a square loss may not be meaningful. An absolute loss function can be used to that $B$ is estimated as
$$\Bhat = \argmin_{B \in St(d,p)} E\{E|Y - E(Y \mid B^T X)| \mid B^TX\}.$$
Clearly, for each distribution of the exponential family, the appropriate loss function should be considered. However, there is a more general loss function that could be considered across all the exponential family distributions: the local deviance function based on local likelihood function. A local deviance has been used for example in \citet{Loader99} for diagnostics purposes.

Regression based on the local log-likelihood evaluated at a given $X \in \Rbb^p$ can be written as
\begin{align*}
L_X(\alpha, \gamma, B) &= \sum_{i=1}^n w_{0i}(X) \log f(Y_i \mid \alpha + \gamma^T B^T (X_i - X)) \\
&= \sum_{i=1}^n w_{0i}(X) \left[\frac{Y_i (\alpha + \gamma^T B^T (X_i - X)) - b(\alpha + \gamma^T B^T (X_i - X))}{a_i(\phi)} + \log f_0(Y_i, \phi)
\right].
\end{align*}
The weights $w_{01}(X), \ldots, w_{0n}(X)$ represent the contribution of each observation toward $L_X(\alpha, \gamma, B)$. Note that the function $a_i(\cdot)$ can vary with $i$ in our formulation; for example, as shown in Appendix~\ref{sec:exp-fams}, Binomial observations with $m_i$ trials will have $a_i(\phi) = 1 / m_i$. However, we assume that $a_i(\phi)$ does not depend on $X$. A deviance of the local likelihood for $Y_j$ at $X_j$ can be expressed as
\begin{align}
D(Y_j, \vartheta(B^T X_j)) = 2 \left[ \max_{\vartheta}\log f(Y_j \mid \vartheta) - L_{X_j}(\alpha_j, \gamma_j, B) \right].
\end{align}
The term $\textstyle{\max_{\vartheta}\log f(Y_j \mid \vartheta)}$ is the maximum local likelihood achievable for an individual observation. The local deviance $D(Y_j, \vartheta(B^T X_j))$ is a measure of the closeness of $Y_j$ to $g^{-1}(\vartheta(B^TX_j))$. Clearly $D(Y_j, \vartheta(B^T X_j))\geq 0$, $D(Y_j, \vartheta(B^T X_j))=0$ if $E(Y_j \mid X_j)=g^{-1}(\vartheta(B^T X_j))$, and $D(Y_j, \vartheta(B^T X_j))$ gets larger when $g^{-1}(\vartheta(B^T X_j))$ gets far from $Y_j$.

Consider minimizing the average deviance $n^{-1} \textstyle{\sum_{j=1}^{n}} D(Y_j, \vartheta(B^T X_j))$ with respect to $(\alpha_j, \gamma_j) \in \Rbb^{d+1}$ for $j = 1, \ldots, n$ and $B \in \Rbb^{p \times d}$ such that $B^T B = I$. This is equivalent to maximizing
\begin{align} \label{eqn:made-glm-obj}
&Q(\alphabf, \gammabf, B)
= \sum_{j=1}^n L_{X_j}(\alpha_j, \gamma_j, B) \\
&\quad= \sum_{j=1}^n \sum_{i=1}^n w_{0i}(X_j)  \left\{
\frac{Y_i (\alpha_j + \gamma_j^T B^T (X_i - X_j)) - b(\alpha_j + \gamma_j^T B^T (X_i - X_j))}{a_i(\phi)} + \log f_0(Y_i, \phi)
\right\}, \nonumber
\end{align}
which is the full local log-likelihood evaluated at each of the sample points, where $\alphabf = (\alpha_1, \ldots, \alpha_n)$ and $\gammabf = (\gamma_1, \ldots, \gamma_n)$. While each sample point $X_j$ has its own regression coefficients $\alpha_j$ and $\gamma_j$, they all share a common dimension reduction kernel matrix $B$. We provide detailed expressions of the MADE objective function $Q$ for several commonly used exponential family outcomes in Appendix~\ref{sec:exp-fams}.

The kernel weights are computed as $w_{0i}(X) =  K_{\Hrm}(X_i-X) /\textstyle{\sum_{j=1}^n} K_{\Hrm}(X_j-X) $ with $K_{\Hrm}(\urm)=|\Hrm|^{-1}K(\Hrm^{-1/2}\urm)$, where $K(u)$ denotes one of the usual multidimensional kernel density functions and the bandwidth $\Hrm$ is a $p \times p$ symmetric and positive definite matrix. For example, the multivariate Gaussian kernel is $K(\urm)= (2 \pi)^{-p/2} \exp\{-\urm^T \urm/2\}$. The choice of the kernel density $K$ and the bandwidth $H$ are next discussed in section~\ref{sec:estimate-coefs}. When the reduction matrix $B$ is known, or an estimator is available, the kernel weights can be refined and written as
\begin{align}
w_{i}(B^TX) = \frac{ K_{\Hrm}((B^T(X_i-X))}{ \sum_{j=1}^n K_{\Hrm}(B^T(X_j-X))}. \label{eq:weight}
\end{align}
These weights now depend the $d$-dimensional data and $\Hrm$ is a $d \times d$ symmetric positive definite matrix.

Note that \eqref{eqn:made-glm-obj} may be naturally extended to the scenario of a training set $\mathcal{M}$ and a test set $\mathcal{N}$ which are not necessarily the same. The test set may contain observations without an observed $Y$ that we wish to predict, or whose $Y$ value has been held out for the purpose of cross-validation. In this case, the MADE objective function becomes
\begin{align} \label{eqn:made-cv-obj}
&Q(\alphabf, \gammabf, B) \\
&\quad= \sum_{j \in \mathcal{N}} \sum_{i \in \mathcal{M}} w_{i}(B^TX_j)  \left\{
\frac{Y_i (\alpha_j + \gamma_j^T B^T (X_i - X_j))
- b(\alpha_j + \gamma_j^T B^T (X_i - X_j))}{a_i(\phi)} + \log f_0(Y_i, \phi)
\right\}. \nonumber
\end{align}
where $\alpha = (\alpha_j : j \in \mathcal{N})$ and $\gamma = (\gamma_j : j \in \mathcal{N})$. We now proceed using \eqref{eqn:made-glm-obj} as the objective function, but note that computations can readily be changed to use \eqref{eqn:made-cv-obj}.

\subsection{Algorithm for Estimation}
\label{sec:estimate-coefs}
The parameters of interest are $\alpha_j, \gamma_j, j=1, \ldots, n$, and $B \in \Rbb^{p \times d}$. We start by assuming that the dimension $d$ is known. For any orthogonal matrix $O$, $\gamma^T B^T = \gamma^T O O^T B^T$, which implies that $\gamma$ and $B$ are not uniquely determined but obtained up to an orthogonal transformation. Furthermore, refined weights based on the Gaussian kernel $w_i(B^T X)$ with $H = hI$ depend on $B$ only through $B B^T = B O O^T B^T$. In this setting, the MADE problem is invariant to orthogonal transformation of $B$ in the sense that
\begin{align*}
Q(\alphabf, \gammabf, B) =
Q(\alphabf, O^T \gamma_1, \ldots, O^T \gamma_n, B O).
\end{align*}
The parameter space of $B$ is the set of $d$-dimensional subspaces in $\Rbb^p$ known as the Grassmann manifold of dimension $d(p-d)$. However, the estimation method we adopt does not estimate all the parameters jointly, but works iteratively. For fixed values of $\alpha_j$ and $\gamma_j$, $j = 1, \ldots, n$, the parameter space of $B$ is the set of $d$-dimensional orthonormal matrices in $\Rbb^p$, also known as Stiefel manifold of dimension $pd - d(d+1)/2$. The dimension of Steifel and Grassmann manifolds is discussed in \citet{EdelmanAS98}. In the following, we present an iterative method to maximize \eqref{eqn:made-glm-obj} for a given dimension $d$, and later discuss selection of $d$.

To estimate the parameters $(\alpha_j, \gamma_j) \in \Rbb^{d+1}, j=1, \ldots, n$, we start by fixing $B$ in \eqref{eqn:made-glm-obj}. We see that maximizing $Q$ over $(\alpha_j, \gamma_j)$ is equivalent to maximizing each $L_{X_j}(\alpha_j, \gamma_j; B)$ separately. There is no closed-form solution of the estimator, except in certain special cases such as Gaussian outcomes. Instead, we proceed with a multivariate Newton-Raphson iterative approach. For a particular $L_{X}(\alpha, \gamma; B)$, let $\xi = (\alpha, \gamma^T)^T$, $Z_i=(1, (X_i-X)^TB)^T$, and $w_i = w_i(B^T X)$ so that
\begin{eqnarray*}
L_X(\alpha, \gamma; B) = \sum_{i=1}^n w_i  \left[
\frac{Y_i \cdot Z_i^T \xi - b(Z_i^T \xi)}{a_i(\phi)} + \log f_0(Y_i, \phi)
\right].
\end{eqnarray*}
Let $Z=(Z_1, \ldots, Z_n)^T$, $W=\diag(w_1, \ldots, w_n)$ and $H(\xi) : \Rbb^{d+1} \rightarrow \Rbb^n$ with entries $[Y_i - b'(Z_i^T \xi)] / a_i(\phi)$ for $i = 1, \ldots, n$. The first derivative at $X$ is then
\begin{equation}
\frac{\partial}{\partial \xi} L_X(\alpha, \gamma; B) = \sum_{i=1}^n w_i \frac{Y_i - b'(Z_i^T \xi)}{a_i(\phi)} Z_i =  Z^T W H(\xi).
\label{eqn:local-score}
\end{equation}
The function $H(\xi)$ has an $n \times (d+1)$ Jacobian
$$J_H(\xi) = \left( \frac{\partial}{\partial \xi_j} \frac{Y_i - b'(Z_i^T \xi)}{a_i(\phi)} \right)
= -\frac{1}{a_i(\phi)} \begin{pmatrix}
b''(z_1^T \xi) Z_{1,1} & \cdots & b''(z_1^T \xi) Z_{1,d+1} \\
\vdots & \ddots & \vdots \\
b''(z_n^T \xi) Z_{n,1} & \cdots & b''(z_n^T \xi) Z_{n,d+1}
\end{pmatrix}.
$$
To formulate Newton-Raphson iterations, suppose $\xi^{(g)}$ is a given iterate and $\xi^{(g)} + \Delta \xi$ will be the next iterate. To solve for $\Delta \xi$ approximately, set the first order Taylor expansion of \eqref{eqn:local-score},
\begin{eqnarray*}
Z^T W H(\xi^{(g)} + \Delta \xi) \approx Z^T W H(\xi^{(g)}) + Z^T W J_H(\xi^{(g)}) \Delta \xi,
\end{eqnarray*}
to zero and solve to obtain $\Delta \xi = -\{ Z^T W J_H(\xi^{(g)}) \}^{-1} Z^T W H(\xi^{(g)}).$
This suggests the update of $\xi$ as $\xi^{(g+1)} = \xi^{(g)} - \{ Z^T W J_H(\xi^{(g)}) \}^{-1} Z^T W H(\xi^{(g)})$.
These steps are iterated until the $g$th iteration where $\lVert \xi^{(g)} - \xi^{(g-1)} \rVert < \varepsilon$ for some small prescribed $\varepsilon > 0$.

To estimate $B$, we suppose that $(\alpha_j, \gamma_j)$, $j=1, \ldots, n$, are fixed and known. Omitting the terms of the objective function \eqref{eqn:made-glm-obj} that are free of $B$, estimation of $B$ is carried out by maximizing
\begin{align}
Q(B) = \sum_{j=1}^n \sum_{i=1}^n w_i(B^T X_j) \frac{1}{a_i(\phi)} \left\{
Y_i (\alpha_j + \gamma_j^T B^T (X_i - X_j)) - b(\alpha_j + \gamma_j^T B^T (X_i - X_j))
\right\},
\label{eqn:B-step}
\end{align}
over the set of $d$ dimensional semi-orthogonal matrices in $\Rbb^p$. In the present work, $B$ is estimated in its natural parameter space, a Stiefel manifold, which naturally honors the orthonormality constraint.

We implemented a conjugate gradient method from \citet{EdelmanAS98} to optimize $Q(B)$ on the Stiefel manifold in the statistical software R \citep{R}. A short background on the algorithm is provided in Appendix~\ref{sec:manifold-opt}. Use of the optimization method requires programming the objective function and its gradient. The gradient of $Q(B)$ is a $p \times d$ matrix with $(r,s)$th entry
\begin{align}
\frac{\partial Q(B)}{\partial B_{rs}} =
\sum_{j=1}^n \sum_{i=1}^n w_{ij}  \Big\{ \frac{Y_i - \mu(\alpha_j + \gamma_j^T B^T (X_i - X_j))}{a_i(\phi)} \Big\}
\gamma_{js} (X_{ir} - X_{jr}),
\end{align}
for $r \in \{1, \ldots p \}$ and $s \in \{1, \ldots, d\}$. Here, $\mu(\vartheta) = b'(\vartheta)$ represents the mean function for the exponential family and $B_{rs}$ represents the $(r,s)$th element of $B$. The optimization on the Stiefel manifold converges when $\tr\{H^TH - (1/2)A^TA\} < \epsilon$ for a user-specified $\epsilon >0$, where $H=\nabla Q(\Bhat)$ and $A=\Bhat^T \nabla Q(\Bhat)$ with $\nabla Q(B)=\partial Q/\partial B - B(\partial Q/\partial B)^T B$.

Joint estimation of all parameters necessitates cycling through the Newton-Raphson iterations for $\alpha_j$ and $\gamma_j$, and the Stiefel manifold optimization for $B$. The procedure is presented as Algorithm~\ref{alg:made}. The full algorithm converges if the estimate of the $B$ matrix (which is common to all observations) converges; this occurs when $\lVert (I - \Bhat_{(r-1)} \Bhat_{(r-1)}^T) \Bhat_{(r)} \rVert_\text{F} < \varepsilon$ for some user-specified $\varepsilon > 0$. Here, $\lVert \cdot \rVert_\text{F}$ is the Frobenius norm and $\Bhat_{(r)}$ is the iterate obtained on the $r$th iteration of the algorithm.
\begin{algorithm}
\caption{MADE algorithm.}
\label{alg:made}
\begin{enumerate}
\item Provide an initial $B$ and weights $w_{ij} = w_i(X_j)$.
\item Do until convergence:
\begin{enumerate}
  \item Fix $B$ and estimate $\alpha_j$ and $\gamma_j$ for $j = 1, \ldots, n$ using Newton-Raphson.
  \item Fix $\alpha$ and $\gamma$ and the weights $w_{ij}$ for $i,j = 1, \ldots, n$, and estimate $B$ using the Stiefel manifold optimization.
  \item Update the weights $w_{ij} = w_i(B^TX_j)$ if refined weights are desired.
\end{enumerate}
\end{enumerate}
\end{algorithm}

A good starting value for $B$ helps for a fast convergence. Practically, any estimator that can be quickly computed can be considered. For example, when dealing with a continuous response, the outer product of gradients method proposed by \citet{Xia02} is one attractive option. Another option is to use the matrix of first $d$ eigenvectors of the fitted covariance matrix $\Sigmahat_{\fit}=\Xbb^T F (F^TF)^{-1}F^T\Xbb/n$ \citep{Cook07}, where $\Xbb$ is the centered $n \times p$ data-matrix of the predictors, and $F$ is a $n \times d$ matrix with $i$th row $(Y_i, Y_i^2, \ldots, Y_i^{d})$.

The choice of kernel density $K(\cdot)$ and bandwidth $\Hrm$ are important in nonparametric regressions, and there is abundant literature on this subject. See for example \citet{FanG92} for continuous-type outcomes and \citet{FanHeckmanWand1995} for exponential family outcomes, as well as the book by \citet{FanGibjels1996}. Optimal bandwidth based on asymptotic mean squared errors is often considered in this literature. \cite{Xia02} instead considered the data-driven cross-validation approach for use with MAVE. Cross-validation selects the bandwidth to minimize an out-of-sample prediction error. In our implementations, we have used the multivariate bandwidth $\Hrm = hI_d$ where $h=cn^{-1/(d+4)}$ for some $c>0$, is the usual optimal bandwidth in the sense of mean integrated squared errors \citep{FanGibjels1996}. 

\section{Inference about $d$}
\label{sec:dim}
The dimension $d$ of the reduction kernel matrix $B$ is to be estimated. Three estimation methods are discussed herein. The first is a sequential permutation test, the second is a bootstrap method, and the third is a cross-validation.

\subsection{Sequential Permutation Test}
\label{sec:perm}
The permutation tests has been used by  \cite{CookW91} to estimate the dimension of sufficient dimension reduction. Their setup was different from ours, but the concept is otherwise identical. The interest is in testing the hypotheses
\begin{equation}
\label{eq:test}
H_0: d=d_0 \text{ vs } H_a: d=d_0+1.
\end{equation}
We propose to test sequentially for $d_0=0, 1, 2, \dots$ until the first time $H_0$ is not rejected. The value $d_0$ is then taken to be the estimated dimension.
Consider testing (\ref{eq:test}) with $d_0=0$. Under the null hypothesis, we have $\vartheta(X_i)=\alpha$ while $\vartheta(X_i)= \alpha + \gamma \beta^T(X_i-X)$ under $H_a$  for any $X_i, i=1, \dots, n$. The parameter $\beta \in \Rbb^p$ resides in a Stiefel manifold so that $\|\beta\|=1$. Thus, testing (\ref{eq:test}) is equivalent to testing $H_0: \gamma=0$ against $H_a: \gamma \neq 0$.

Now consider testing (\ref{eq:test}) with $d_0>0$. Let $B_0 \in \Rbb^{p \times d_0}$ and $B \in \Rbb^{p \times (d_0+1)}$ be the reduction kernel matrices under $H_0$ and $H_a$, respectively.  We can then write $B=[B_0, \beta]$ where $\beta \in \Rbb^p$ such that $\beta^T B_0=0$ and $\|\beta\|=1$. The canonical parameter can be written as $\vartheta(X_i) = \alpha + \gamma_0^T B_0^T(X_i-X)$ under $H_0$ while $ \vartheta(X_i) = \alpha + \gamma_0^T B_0^T(X_i-X) + \gamma \beta^T(X_i-X)$ under $H_a$, at any $X$ where $\gamma_0 \in \Rbb^{d_0}$ and $\gamma \in \Rbb$. Again, testing $d=d_0$ against $d=d_0+1$ is equivalent to testing $\gamma=0$ against $\gamma \neq 0$.

The sample value of $\gamma$ and its distribution are needed to carry out the test. However, for a given data set with $n$ observations, there are $n$ local parameters $\gamma_j, j=1, \dots, n$ to estimate. Under $H_0$, we expect each of these $n$ estimates $\gammahat_j$ close to zero. Let $u=\rho(\gammahat_{1}, \dots, \gammahat_{n})$ be a summary statistic of these $n$ estimates using the unperturbed $Y$ values. For example, $\rho(.)$ can be the sample mean or other summary statistics.

To obtain a sampling distribution of these estimates under $H_0$, generate a large number $R_\text{perm}$ of permutations $Y^{(r)}=(Y_1^{(r)}, \ldots, Y_n^{(r)})$ of the $Y$-observations, which yields $\gammahat_{jr}$ for $j=1, \dots, n,$ and $r=1, \dots, R_\text{perm}$. Denote $\hat{u}_r = \rho(\gammahat_{1}^{(r)}, \dots, \gammahat_{n}^{(r)})$ as the summary obtained from $Y^{(r)}$ for $r=1, \dots, R_\text{perm}$. We would expect $\hat{u}$ to be `close' to $\hat{u}_r$ if $H_0$ is true, and far from $\hat{u}_r$ otherwise. The fraction
$$\frac{1}{R_\text{perm}}\sum_{r=1}^{R_\text{perm}} I(|\hat{u}_{r}| > |\hat{u}|)$$
is an approximate p-value to test \eqref{eq:test}. The procedure is summarized as follows.
\begin{itemize}
	\item[] Starting with $d_0=0$, do until $H_0: d=d_0$ is not rejected
		\begin{enumerate}
			\item Estimate $\hat{u} = \rho(\gammahat_1, \dots, \gammahat_n)$
			\item Generate $Y^{(r)} = (Y_1^{(r)}, \dots, Y_n^{(r)})$ and compute $\hat{u}_r = \rho(\gammahat_1^{(r)}, \dots, \gammahat_n^{(r)})$, $r=1, \dots, R_\text{perm}$
			\item If $H_0$ if rejected then $d_0 = d_0+1$.
		\end{enumerate}
\end{itemize}
Practically, for testing with $d_0=0$, all the parameters are estimated directly using the estimation method in section~\ref{sec:estimate-coefs}. However, for $d_0>0$, we first obtained $\Bhat_0$, an estimate of $B_0$ under $H_0$. Then $\alpha, \gamma_0, \gamma$, and $\beta$ are estimated with $B_0$ replaced by $\Bhat_0$. The sample mean is used in our implementation for the summary $\rho(.)$.

\subsection{A Bootstrap Method}
\label{sec:boot}
We propose bootstrapping a statistic similar to the likelihood ratio test (LRT) statistic to test the dimension. Consider again testing \eqref{eq:test} sequentially. Let $\Bhat_0$ be the MADE estimate under $d = d_0$, which yields the fitted local coefficients $(\hat{\alpha}_{0j}, \hat{\gamma}_{0j})$ for $j = 1, \ldots, n$, and hence the fitted regressions $\widehat\vartheta_0(X_j)$ for $j = 1, \ldots, n$. Similarly, denote $\Bhat$ as the MADE estimate under $d = d_0 + 1$, which yields fitted local coefficients $(\hat{\alpha}_{j}, \hat{\gamma}_{j})$ and fitted regressions $\widehat\vartheta(X_j)$ for $j = 1, \ldots, n$. The fitted regressions from MADE may be used to evaluate the likelihood based on \eqref{eqn:exp-fam}, which is
\begin{align*}
L(\vartheta(X_1), \ldots, \vartheta(X_n)) = \prod_{i=1}^n
\exp\left\{ \frac{Y_i \vartheta(X_i) - b(\vartheta(X_i))}{a_i(\phi)}  \right\} f_0(Y_i, \phi).
\end{align*}
A quantity analogous to the LRT statistic may be computed as
\begin{align*}
\lambdahat = 2 \left[ \log L(\widehat\vartheta(X_1), \ldots, \widehat\vartheta(X_n)) - \log L(\widehat\vartheta_0(X_1), \ldots, \widehat\vartheta_0(X_n)) \right].
\end{align*}
We do not know the distribution of $\lambdahat$ under the null hypothesis. To fully specify the test procedure, a parametric bootstrap procedure can approximate the null distribution from the data.
\begin{itemize}
	\item[] Starting with $d_0=0$, do until $H_0: d=d_0$ is not rejected
	\begin{enumerate}
		\item Obtain $\lambdahat$, an estimate of $\lambda$ using the original sample.
		\item Draw a bootstrap sample $Y^{(r)} = (Y_1^{(r)}, \ldots, Y_n^{(r)})$ from the null likelihood $L(\hat\vartheta_0(X_1), \ldots, \hat\vartheta_0(X_n))$.
		\item  Estimate $\widehat\vartheta^{(r)}_0(X_1), \ldots, \widehat\vartheta^{(r)}_0(X_n)$ and $\widehat\vartheta^{(r)}(X_1), \ldots, \widehat\vartheta^{(r)}(X_n)$ under $H_0$ and $H_1$ respectively using MADE with data $\{(Y_i^{(r)}, X_i) : i = 1, \ldots, n\}$.
		\item Compute $\lambdahat^{(r)} = 2 [\log L(\hat\vartheta^{(r)}(X_1), \ldots, \hat\vartheta^{(r)}(X_n)) - \log L(\hat\vartheta^{(r)}_0(X_1), \ldots, \hat\vartheta^{(r)}_0(X_n)) ]$.
		\item Repeat steps 2--4 for $r = 1, \ldots, R_\text{boot}$, where $R_\text{boot}$ is the desired number of bootstrap iterations.
		\item If $H_0$ if rejected then $d_0=d_0+1$.
	\end{enumerate}
\end{itemize}
An approximate p-value can now be computed as
\begin{align*}
\frac{1}{R_\text{boot}} \sum_{r=1}^{R_\text{boot}} I(\lambdahat^{(r)} \geq \lambdahat).
\end{align*}

\subsection{Cross-validation}
In the context of local likelihood approach, \cite{Xia02} considered a cross-validation in conjunction with a prediction method based on Nadaraya-Watson kernel. We provide a similar approach with an alternative prediction method.

The true dimension $d$ of $B$ can be estimated to yield the best predictive model on out-of-sample observations. We propose a $K$-fold cross-validation to estimate the mean squared prediction error \citep{HastieTF09}. Suppose $D = \{ 1, \ldots, n \}$ contains all indexes in the dataset. Let us partition $D$ randomly into $K$ subsets $D_1, \dots, D_K$ of approximately equal sizes, and let $D_{-k}$ be the subset of $D$ where $D_k$ is held out. Denote $\hat{Y}_{j,-k}^{(d)}$ the predicted value for observation $j \in D_k$, where $D_k$ is a test set for evaluation and $D_{-k}$ is a training set. The parameters are estimated for a fixed dimension $d$ using $D_{-k}$. Taking $\Lcal(Y,\hat{Y})$ as a predetermined loss function, we estimate the dimension $d$ as
$$\dhat= \argmin_{d \in \{0,1,2, \dots, p\}} \sum_{k=1}^K \sum_{j \in D_k} \Lcal(Y_j, \hat{Y}_{j,-k}^{(d)}).$$
The prediction values $\hat{Y}_{j,-k}^{(d)}$ are computed according to one of the three approaches outlined in section~\ref{sec:pred}. For example, the loss $\Lcal(Y,\hat{Y})$ may be a squared loss for continuous responses, or an absolute loss for a Bernoulli outcome.

\subsection{Connection to MAVE}
Minimum average variance estimation, or MAVE was proposed by \citet{Xia02}. It is an adaptive estimation method using a local estimation to determine a dimension reduction of $X$ in the regression of $Y|X$. It assumes that a model of $Y|X$ is of the form $Y=m(B_0^TX) + \epsilon$, where $m$ is an unknown smooth link function, and $B_0$ is a $p \times d$ semi-orthogonal matrix so that $B_0^T B_0=I_d$. There is no extraneous distributional assumption, however, in that formulation, $Y|X$ has the same distribution as $Y|B_0^TX$. The direction $B_0$ was then determined as the solution of
$$ \min_{B: B^TB=I} E\{[E\{Y-E(Y|B^TX)^2\}|B^TX]\}$$
Using the approximation that $g(B^TX_i) \approx a_j + b_j^T B^T(X_i-X_j)$ at any $X_j$, let $\abf=\{a_1, \cdots, a_n\}$, and $\bbf=\{b_1, \cdots, b_n\}$, and let the weights $w_{i}(B^TX_j)$ as in expression~(\ref{eq:weight}). The local parameters in $\abf$ and $\bbf$, and the matrix $B$ are estimated essentially as
\begin{align*}
\{\abfhat, \bbfhat, \Bhat\}= \argmin_{{\abf, \bbf, B: B^TB=I}} \sum_{j=1}^n \sum_{i=1}^n
\left\{ Y_i - (a_j + b_j^T B^T (X_i - X_j)) \right\}^2 w_{i}(B^TX_j).
\end{align*}

Now let write the local deviance version in the case of Gaussian outcome $Y$ where the variance $\sigma^2$ is assumed fixed and known. The local parameters $\alphabf = \{\alpha_1, \cdots, \alpha_n\}$ and $\gammabf=\{\gamma_1, \cdots, \gamma_n\}$, and the parameter $B$ are estimated as
\begin{eqnarray*}
\{\alphabfhat, \gammabfhat, \Bhat\} &=& \argmax_{\alphabf, \gammabf, B}
\sum_{j=1}^n \sum_{i=1}^n \left\{-\frac{1}{2} \log 2 \pi \sigma^2 -\frac{ [Y_i - (\alpha_j + \gamma_j^T B^T (X_i - X_j))]^2 }{2\sigma^{2}}\right\} w_{i}(B^TX_j) \\
&=& \argmin_{\alphabf, \gammabf, B} \sum_{j=1}^n \sum_{i=1}^n \left\{ Y_i - (\alpha_j + \gamma_j^T B^T (X_i - X_j)) \right\}^2 w_{i}(B^TX_j) .
\end{eqnarray*}
Clearly MADE with Gaussian outcomes is equivalent to MAVE, thus MADE effectively subsumes MAVE. In the Gaussian case of MADE, the normality assumption does not add any limitation in the formulation nor in the estimation. It is noteworthy that in the setting of MAVE, \cite{Xia02} an iterative least squares method is employed to estimate $\alphabf$ and $\gammabf$. Furthermore, a quadratic programming method was used. The orthogonality constraint of $B$ was dealt with by estimating individual columns of $B$ sequentially and orthonormalizing these column-vectors to obtain $\Bhat$. In our case, we broke away from the procedure of \cite{Xia02} by carrying out the estimation of $B$ on its natural space, which is a Stiefel manifold.

\section{MADE with Two-Parameters Exponential Family}
\label{sec:twopar}

Discuss the case where both $\phi = \phi(X)$

For non-Gaussian exponential family types, the dispersion parameter $\phi$ may not simply cancel out of the MADE objective function. When $\phi$ is not known, it must be estimated within MADE; this can be accomplished by adding a step to Algorithm~\ref{alg:made}, which is given as Algorithm~\ref{alg:made-dispersion}.

Here, the objective function is free of the dispersion parameter $\phi = \sigma^2$, and an unknown $\sigma^2$ can be estimated outside of MADE. For example, given estimates $\hat{\mu}_1, \ldots, \hat{\mu}_n$ computed from MADE, we may consider the likelihood of observations $Y_i \stackrel{\text{ind}}{\sim} \text{N}(\hat{\mu}_i, \sigma^2)$ and accordingly use the maximum likelihood estimator $\hat{\sigma}^2 = n^{-1} \textstyle{\sum_{i=1}^n} (Y_i - \hat{\mu}_i)^2$. For non-Gaussian exponential family types, the dispersion parameter $\phi$ may not simply cancel out of the MADE objective function. When $\phi$ is not known, it must be estimated within MADE; this can be accomplished by adding a step to Algorithm~\ref{alg:made}, which is given as Algorithm~\ref{alg:made-dispersion}. We will proceed using refined weights for the remainder of the paper, and will make use only of Algorithm~\ref{alg:made}.

We will proceed using refined weights for the remainder of the paper, and will make use only of Algorithm~\ref{alg:made}.

\begin{algorithm}
\caption{MADE algorithm with unknown dispersion parameter.}
\label{alg:made-dispersion}
\begin{enumerate}
\item Provide an initial $B$ and weights $w_{ij} = w_i(X_j)$.
\item Do until convergence:
\begin{enumerate}
  \item Fix $(B, \phi)$ and estimate $\alpha_j$ and $\gamma_j$ for $j = 1, \ldots, n$ using Newton-Raphson.
  \item Fix $\phi$, $\alpha$ and $\gamma$ and the weights $w_{ij}$ for $i,j = 1, \ldots, n$, and estimate $B$ using the Stiefel manifold optimization.
  \item Fix $(B, \alpha, \gamma)$ and estimate $\phi$ by maximizing the resulting objective function.
  \item Update the weights $w_{ij} = w_i(B^T X_j)$ if refined weights are desired.
\end{enumerate}
\end{enumerate}
\end{algorithm}

\section{Simulations}
\label{sec:sim}
\subsection{Estimation of $B$}
\label{sec:sim_na}

We study the performance of MADE in estimating the reduction subspace under several settings for the distribution of response $Y$. An empirical consistency of the estimate $\Bhat$ is evaluated together with a contrast to MAVE of \cite{Xia02} and PFC of \cite{Cook07}. We report the results for Binomial, Gaussian, and Poisson distributions. Under each setup, a dataset was generated with a specified matrix $B \in \Rbb^{p \times d}$ representing the subspace $\Scal_B$. The MADE, MAVE, and PFC methods are then used to obtain $\Scal_{\Bhat}$, the estimator of $\Scal_{B}$, where the dimension of $\Bhat$ was not estimated but set to the true $d$. To compare $\Scal_{B}$ to $\Scal_{\Bhat}$, we used the distance $\rho(\Scal_{B}, \Scal_{\Bhat}) = \|(I-\Bhat \Bhat^T)B\|$ suggested in \cite{Xia02}.
For a given sample size $n$, the procedure was repeated one hundred times. Following is the description of the data generation under the three aforementioned setups. In all cases, $\beta=(-1,1,-1,2,-2,2)^T/\sqrt{15}$ and $n$ was varied from 25 to 400.
\begin{enumerate}
\item \textit{Binomial:} We first generated the response vector $\Ybb$ as $n$ independent $\text{Bernoulli}(0.7)$. Then the predictors were generated as $\Xbb^T =\beta\Ybb^T + \sigma \ebb$, $\sigma=0.5$ and the elements of $\ebb$ are independently sampled from a standard normal distribution.

\item \textit{Gaussian:} The response was obtained as $Y \sim \text{N}\big( \exp(1.8 \beta^T X) / (1+\exp(5[\beta^T X]^2)), 0.3^2 \big)$ with $X =(V_{1}, V_{2}, V_{3}, V_{4}, V_{5}, V_{6})^T$. The predictors were generated as $V_{1} \sim \text{Bernoulli}(0.7)$, $V_{2} \sim \text{Binomial}(5, 0.8)$, $V_{3} \sim \text{Exponential}(3)$, $V_{4} \sim \text{Exponential}(3)$, $V_{5} \sim \text{Uniform}(-2, 2)$, $V_{6} \sim \text{Gamma}(5, 10)$.

\item \textit{Poisson:} The response was generated as $Y \sim \text{Poisson}(3.5 \exp(\sin(\pi \beta^T X/2)))$, and the six predictors were obtained from $\text{Uniform}(0, 3)$.
\end{enumerate}
\begin{figure}[!ht]
\centering{
\begin{tabular}{ccc}
\includegraphics[width=50mm, height=50mm]{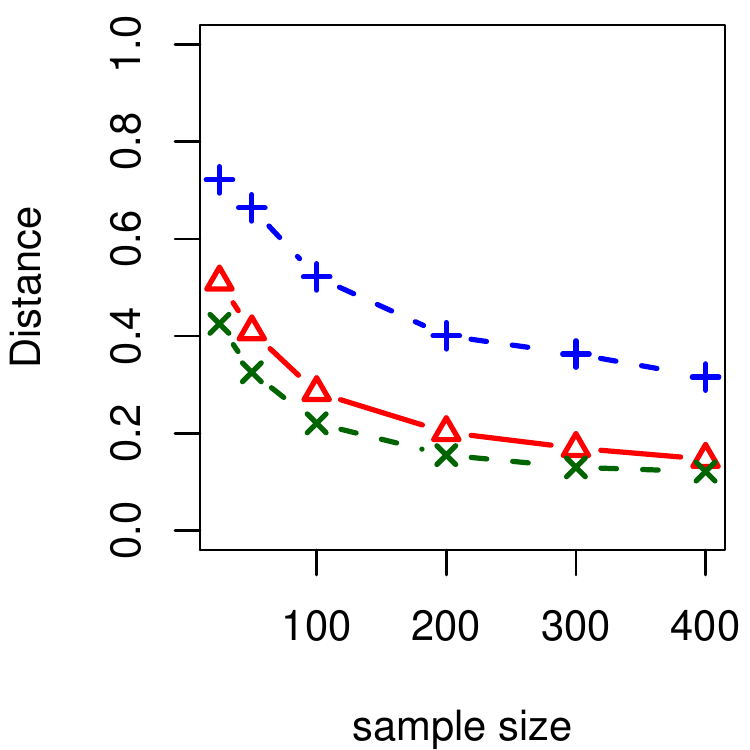} &
\includegraphics[width=50mm, height=50mm]{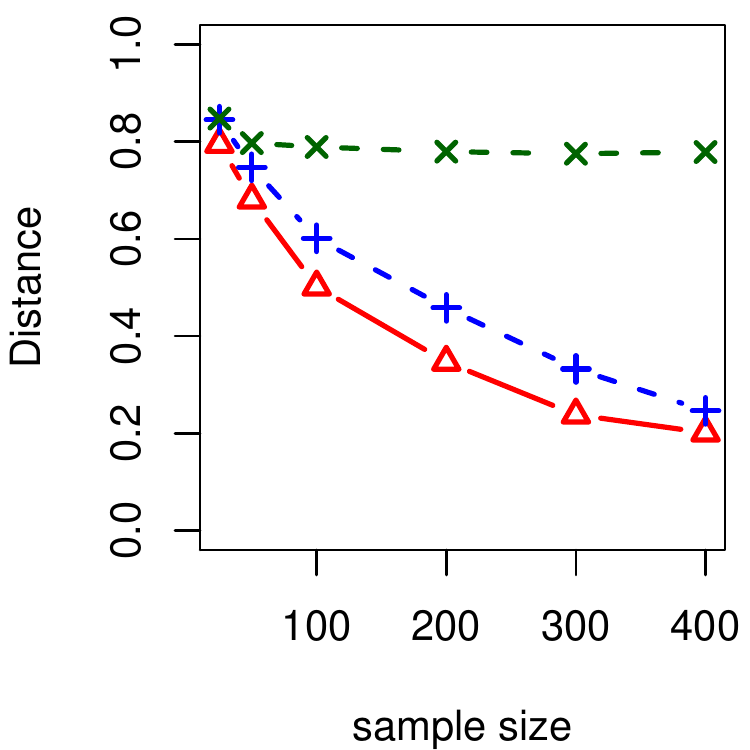} &
\includegraphics[width=50mm, height=50mm]{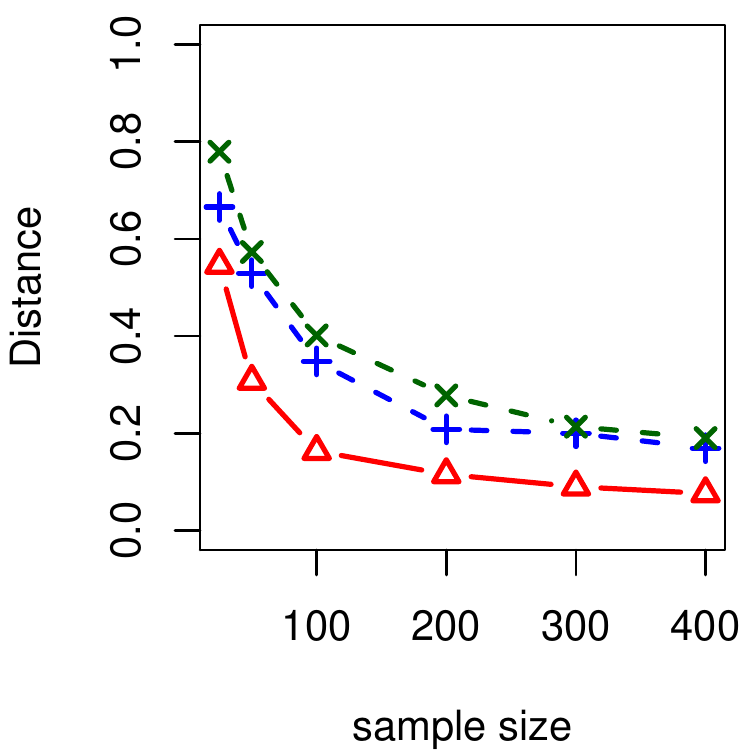}\\
(a) Binomial & (b) Gaussian & (c) Poisson
\end{tabular}
\caption{Distance $\rho(\Scal_{B}, \Scal_{\Bhat})$ with MADE ({\color{red} ``$\triangle$"}), MAVE ({\color{blue} ``$+$"}), and PFC ({\color{darkgreen} ``$\times$"}).}\label{Fig:na}}
\end{figure}
Figures~\ref{Fig:na}a-c show distances $\rho(\Scal_{B}, \Scal_{\Bhat})$ for varying values of $n$. Overall, it appears that the MADE estimator converges to the true $B$ as the sample size increases. For the Binomial case, PFC dominated MADE, and MAVE showed the worse performance. In the Gaussian and Poisson cases, MADE outperformed MAVE, and both dominated PFC. It is possible that the poor performance of PFC in the latter two cases is due to the data generation scheme, which is based on forward regression.

\subsection{Prediction}
\label{sec:pred}
Suppose we wish to estimate $E(Y \mid X)$ for a new observation $X = X^{*}$. Let $\Bhat$ denote the estimate of $B$ based on $n$ independent observations. We provide three different prediction methods that do not rely on the exact specification of the regression function to predict the response corresponding to a new observation $X^{*}$. Let $\{w_{i*}\}_{i=1}^{n}$ denote the set of kernel weights obtained as
$w_{i*} = \textstyle{K_{\Hrm}(\Bhat^T(X_i-X_{*}))/\sum_{m=1}^n K_{\Hrm}(\Bhat^T(X_m-X_{*}))}$. The first prediction method yields the predicted response as
\begin{equation} \label{kernelweight}
\Ehat(Y \mid X^{*}) = \sum_{i=1}^{n} w_{i*} Y_i =
\frac{\sum_{i=1}^{n} K_H(\Bhat^T(X_i-X_{*}))Y_i}{\sum_{i=1}^{n} K_H(\Bhat^T(X_i-X_{*}))}.
\end{equation}
This prediction method is essentially a Nadaraya-Watson estimator which is typical for nonparametric methods, and was used in \cite{Xia02} in the context of cross-validation. It can be used with any dimension reduction method that could provide an estimate for $B$. We will refer to this prediction method as the NW method.

The second prediction method is relative to local likelihood regression. We continue to assume that the $n$ independent samples are used to obtain $\Bhat$, an estimator of the reduction matrix $B$. For the new observation $X^{*}$, $\alphahat$ and $\gammahat$ may be are obtained as
\begin{align*}
(\alphahat, \gammahat) &= \argmax_{\alpha, \gamma} \sum_{i=1}^n w_{i}(\Bhat^T X^*)  \Big\{
\frac{Y_i (\alpha + \gamma^T \Bhat^T (X_i - X^*)) - b(\alpha + \gamma^T \Bhat^T (X_i - X^*))}{a_i(\phi)} \Big\},
\end{align*}
Recall in the original Taylor expansion that $\alpha = \vartheta(B^T X^*)$; as is often done in local likelihood literature \citep{FanG92, Loader99} we may predict $Y^*$ using only the intercept as $g^{-1}(\alphahat)$. We will denote this as local likelihood prediction I, or $\Ehat(Y \mid X^{*})_{\text{LL}_\text{I}}$.

We also consider local likelihood prediction II, computed as
$$\Ehat(Y \mid X^{*})_{\text{LL}_\text{II}} = g^{-1}\left( \sum_{i=1}^n (\alphahat + \gammahat^T \Bhat^T(X_i-X^{*}))w_{i*} \right),$$
which incorporates the estimate for the slope as well. We will now compare these three prediction methods in a simulation study. The datasets were generated as in section~\ref{sec:sim_na} except for the following details.
\begin{enumerate}
\item \textit{Binomial}: the response observations were obtained from $\text{Bernoulli}(0.52)$.
\item \textit{Gaussian}: the observations were generated as $Y \sim \text{Normal}\big( e^{1.8 \beta^T X}/(1+e^{5(\beta^T X)^2}), 0.3^2 \big)$.
\item \textit{Poisson}: the elements of $X$ were obtained from the $\text{Uniform}(-1, 2)$, and the response was generated as  $Y \sim \text{Poisson}(4 e^{(\sin[\pi \beta^T X/2])})$.
\end{enumerate}
%

For each step of the simulation, a training set of $n$ observations was generated for estimation of the parameters, and an additional test set of $n_e$ observations was generated to predict the response.
\begin{figure}[!ht]
\centering
\begin{tabular}{ccc}
\includegraphics[width=50mm, height=50mm]{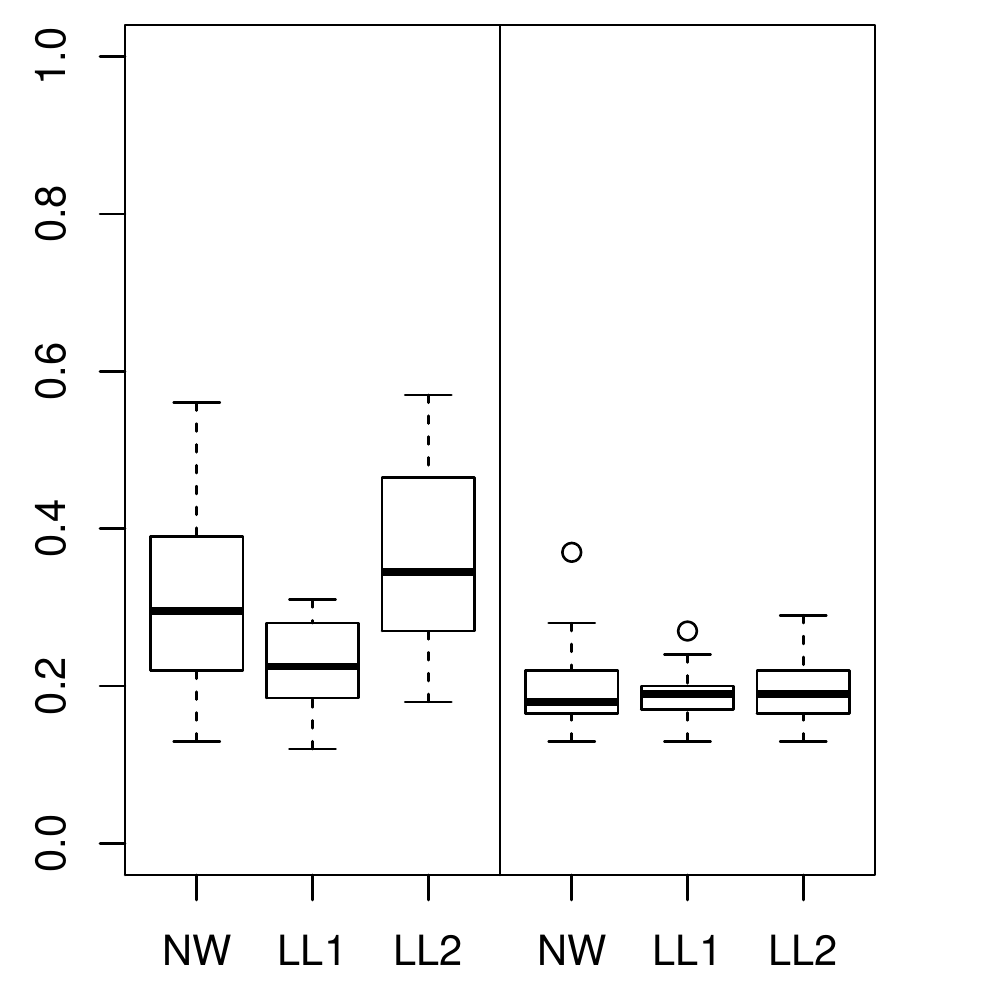} &
\includegraphics[width=50mm, height=50mm]{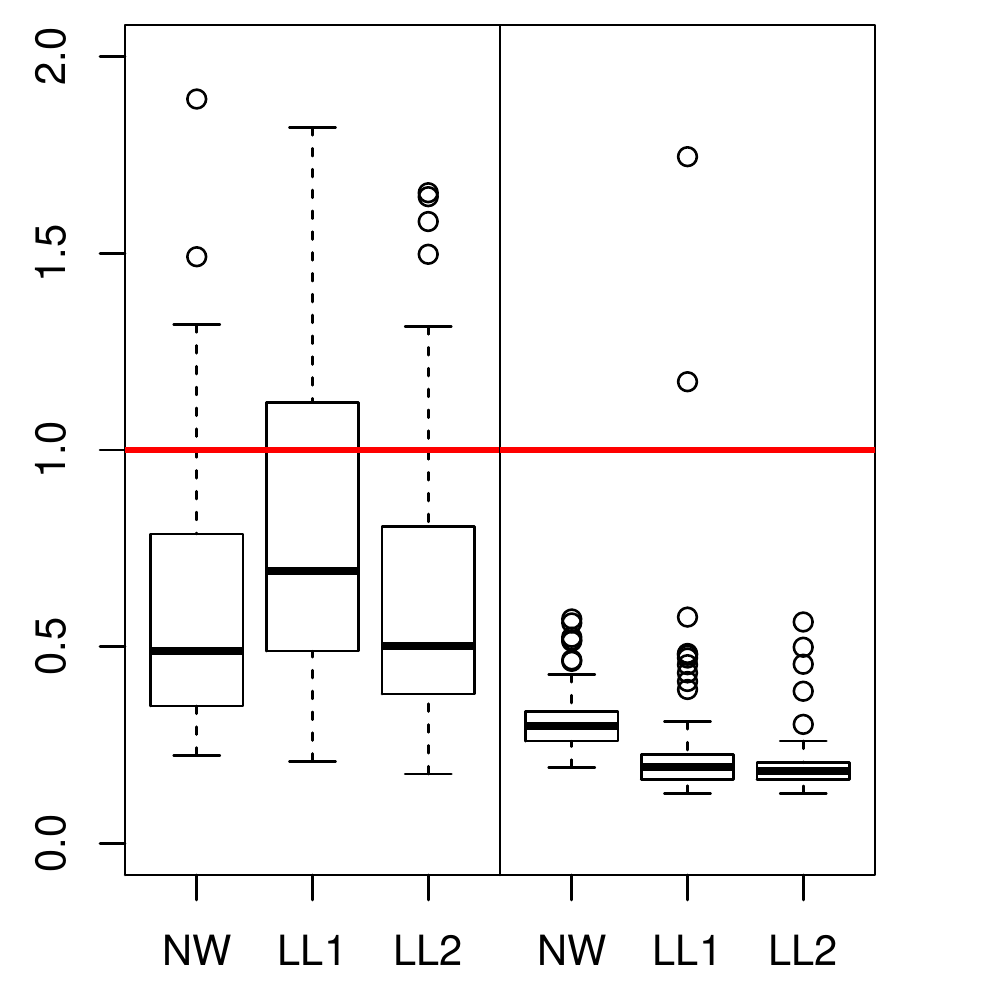} &
\includegraphics[width=50mm, height=50mm]{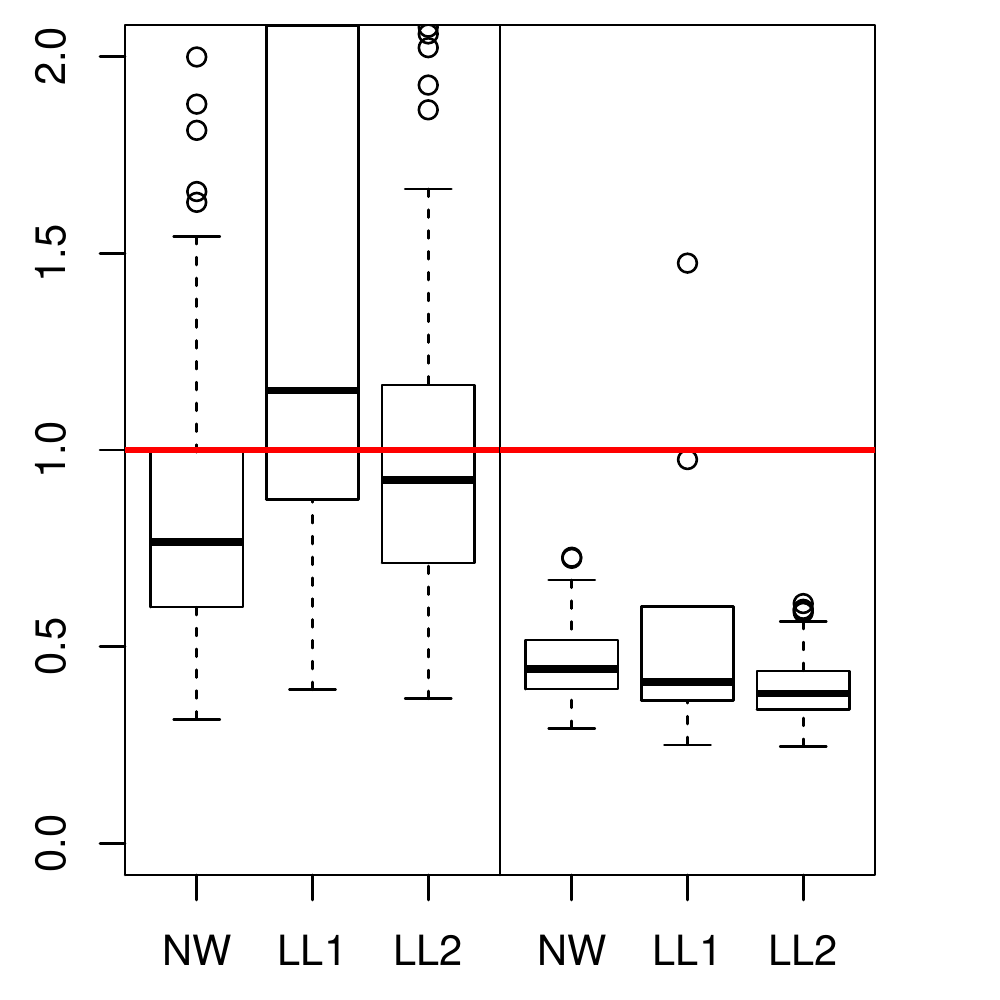} \\
(a) Binomial & (b) Gaussian  & (c) Poisson
\end{tabular}
\caption{Prediction error using NW, LL1, and LL2. For each plot, the first three boxplots are with $n=20$, the next three are for with $n=200$.
\label{Fig:pred}}
\end{figure}
The prediction errors were calculated as follows. For the Binomial case where the observations are binary, the prediction error was a misclassification error obtained as $n_e^{-1} \textstyle{\sum_{i=1}^{n_e}} I(Y_i \neq \Yhat_i)$. For both Gaussian and Poisson, we used the usual mean squared prediction error $n_e^{-1} \textstyle{\sum_{i=1}^{n_e} (Y_i-\Yhat_i)^2}$. We note that for a Bernoulli response, NW estimates the probability $\Ehat(Y \mid X) = \hat{P}(Y=1 \mid X)$ so that $\Yhat=1$ if $\hat{P}(Y=1 \mid X) \geq 0.5$. The prediction errors were averaged over one hundred replications. This process was repeated for $n=20$ and also for $n=200$, while $n_e$ was fixed to 200.

Figures~\ref{Fig:pred} shows the results. Mean squared prediction errors in both Gaussian and Poisson cases were scaled to the variance of $Y$, thus these prediction errors (as well as the misclassification error) were expected to be between 0 and 1. Overall, none of the prediction methods appears to be best. We notice for the binary outcome in Figure~\ref{Fig:pred}(a) that LLI performed slightly better and showed less variability than the other two, for both small and large sample sizes. However, for the Gaussian outcome in Figure~\ref{Fig:pred}(b) and Poisson outcomes in Figure~\ref{Fig:pred}(c), LLI showed unexpected larger variabilities than the other two methods for smaller sample size. But all three methods performed comparably with larger sample sizes. It should be mentioned that the bandwidth was hand-picked once and kept constant over all simulations.

\section{Applications}
\label{sec:appl}

We present three data analyses to illustrate the use of MADE. The first illustration uses the 'flea' dataset \citep{Lubischew62} where the response is categorical with three levels. The second is 'Big Mac' dataset \citep{Enz91} with a continuous response, and the third example is with 'fishing' dataset \citep{Count14} with a count response. In all three cases, we have used a Gaussian kernel density with a hand-picked bandwidth.

\subsection{Flea Data}

The flea dataset \citep{Lubischew62} contains 74 observations on six variables regarding three species of flea-beetles: \textit{concinna, heptapotamica}, and \textit{heikertingeri}. The species are taken as a categorical response, while measurements on the six remaining variables are continuous. These six variables are \textit{tars1}, the width of the first joint of the first tarsus in microns, \textit{tars2}, the same for the second joint, \textit{head}, the maximal width of the head between the external edges of the eyes in 0.01 mm, \textit{aede1}, the maximal width of the aedeagus in the fore-part in microns, \textit{aede2}, the front angle of the aedeagus (1 unit = 7.5 degrees), and \textit{aede3}, the aedeagus width from the side in microns.

The goal is to find the sufficient directions that best separate the three species. A multinomial response with three levels is most appropriate for this dataset. However, Multinomial-MADE would require support for multivariate canonical parameters, which has yet not been developed. To circumvent this issue, we consider two Binomial fits. We considered one class, \textit{concinna}, to be the `success' class. The six covariates were centered and scaled to have unit variance. In the first Binomial-MADE fit, the data are taken to be only the \textit{concinna} and  \textit{heptapotamica} classes. The direction $B_1 \in \Rbb^{6}$ that best separates the two classes is then estimated. In the second fit, the data is taken to be just the \textit{concinna} and \textit{heikertingeri} classes. The direction $B_2 \in \Rbb^{6}$ that best separates the two classes is also estimated. For each of the two Binomial-MADE models, the bandwidths were set using optimal bandwidth with $c=1, d=1$, and $n=43$ and 53, respectively for the first and second fits.

Figure~\ref{Fig:fleamac}(a) provides the plot of the two sufficient dimension reduction components $\Bhat_1^T X$ and $\Bhat_2^T X$, and the estimates for the two reductions are given in Table~\ref{Tab:flea-est}. Clearly, these two components allow a separation of the three species that is noticeable graphically. We notice that \textit{tars1} and \textit{aede2} are the most prominent terms in the first component $\Bhat_1$, and (\textit{aede1, aede3}) are the most dominant variables in the second component.

\begin{table}[b]
\centering
\caption{Estimated reduction directions for flea data.}
\label{Tab:flea-est}
\tt
\begin{tabular}{rrrrrrr}
&
tars1 &
tars2 &
head &
aede1 &
aede2 &
aede3 \\
\hline
$\Bhat_1$ & -0.588 & -0.206 &  0.049 & -0.295 & -0.721 &  0.047 \\
$\Bhat_2$ &  0.301 & -0.314 & -0.091 & -0.691 & -0.045 & -0.568 \\
\hline
\end{tabular}
\end{table}

\subsection{Big Mac Data}

The Big Mac dataset \citep{Enz91} gives the average values in 1991 on several economic indicators for 45 world cities, and contains ten continuous variables $X$. The response $Y$, which is continuous, is the minimum labor to purchase one Big Mac in US dollars. The interest is in regression of $Y$ on $X$, specifically a sufficient reduction of $X$ such that $Y \indep X \mid B^TX$, with $B \in \Rbb^{10 \times d}$ where $d$ is to be estimated. The predictors are \textit{Bread, BusFare, EngSal, EngTax, Service, TeachSal, TeachTax, VacDays}, and  \textit{WorkHrs}, respectively, minimum labor to buy a BigMac and fries, minimum labor to buy one kilogram of bread, lowest cost of 10 kilometers ride on public transit, electrical engineer annual salary, tax rate paid by engineer, annual cost of 19 services, primary teacher salary, tax rate paid by primary teacher, average days vacation per year, average hours worked per year.

\begin{figure}[!h]
\centering
\begin{tabular}{cc}
\includegraphics[width=0.3\textwidth]{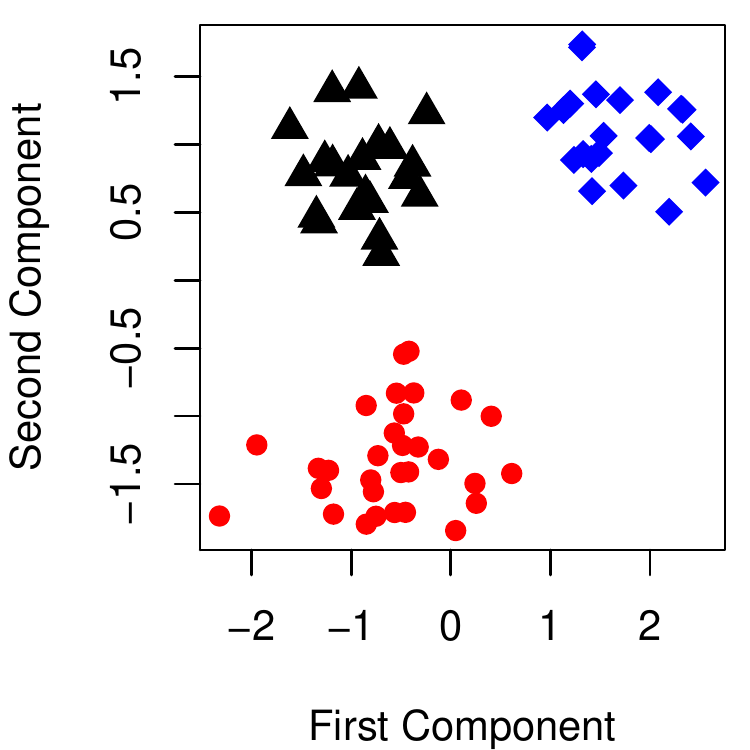} &
\includegraphics[width=0.3\textwidth]{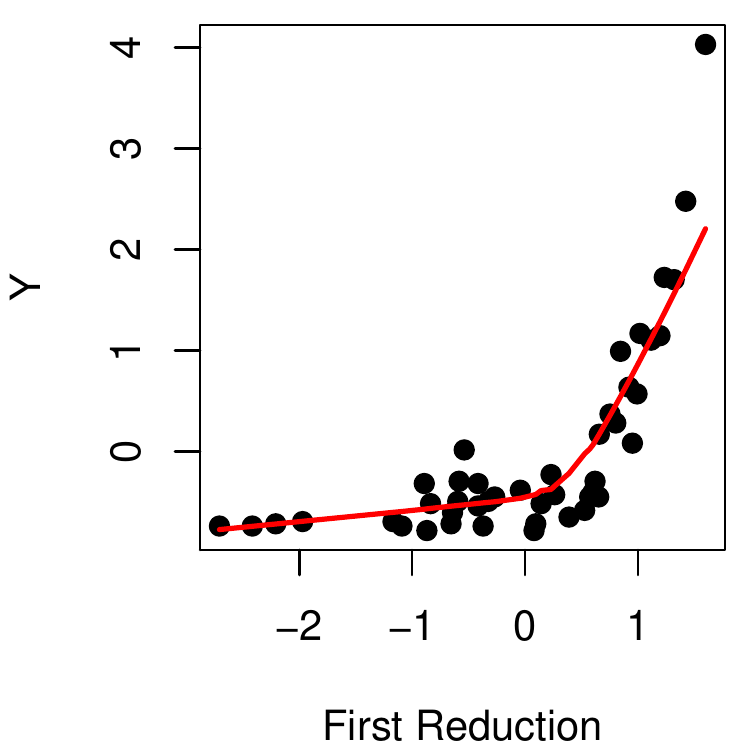} \\
(a) Flea & (b) Big Mac
\end{tabular}
\caption{Plots of the two sufficient reductions for the flea dataset (a) and of the response against the single-index reduction $\Bhat^T X$ for the Big Mac dataset (b).}
\label{Fig:fleamac}
\end{figure}

We centered and scaled $X$ to have mean zero and unit variance prior to fitting MADE. We sequentially tested the hypothesis $H_0: d=d_0$ against $H_a: d > d_0$, starting with $d_0=0$. Using the permutation test described in section~\ref{sec:perm}, 100 permutations were generated. We tested at 0.05 significance level and rejected the null hypothesis for $d_0=0$, but failed to reject for $d_0=1$. Consequently, the sufficient reduction of $X$ is unidimensional. The same estimated dimension was obtained using the bootstrap method described in \ref{sec:boot} with $R_\text{boot}=100$. For the bandwidth, we have taken $h=0.47$ (optimal bandwidth with $c=1, n=45, d=1$) Figure~\ref{Fig:fleamac}(b) shows a nonlinear relationship between the response against $\Bhat^T X$, which is expected from an initial graphical exploratory data analysis. The estimate for the reduction kernel matrix is given in Table~\ref{Tab:bigmac-est}. Variables \textit{EngSal} seems the most prominent variable while \textit{Bread} is the least in explaining the response.

\begin{table}[b]
\centering
\small
\caption{Estimated reduction direction for Big Mac data.}
\label{Tab:bigmac-est}
\tt
\begin{tabular}{rrrrrrrrrr}
&
Bread &
BusFare &
EngSal &
EngTax &
Service &
TeachSal &
TeachTax &
VacDays &
WorkHrs \\
\hline
$\Bhat$ & 0.012 & -0.091 & -0.876 & 0.158 & 0.044 & -0.319 & 0.173 & -0.246 & -0.071 \\
\hline
\end{tabular}
\end{table}

\subsection{Fishing Data}
The fishing dataset \citep{Count14} has 147 observations and only three continuous predictors. The response is \textit{totabund}, the total fish counted per site. The predictors are
\textit{density, log(meandepth)}, and \textit{log(sweptarea)}, respectively, folage density index, the natural logarithm of mean water depth per site, and the natural logarithm of the adjusted area of site.

\begin{figure}[!h]
\centering
\begin{tabular}{cc}
\includegraphics[width=0.3\textwidth]{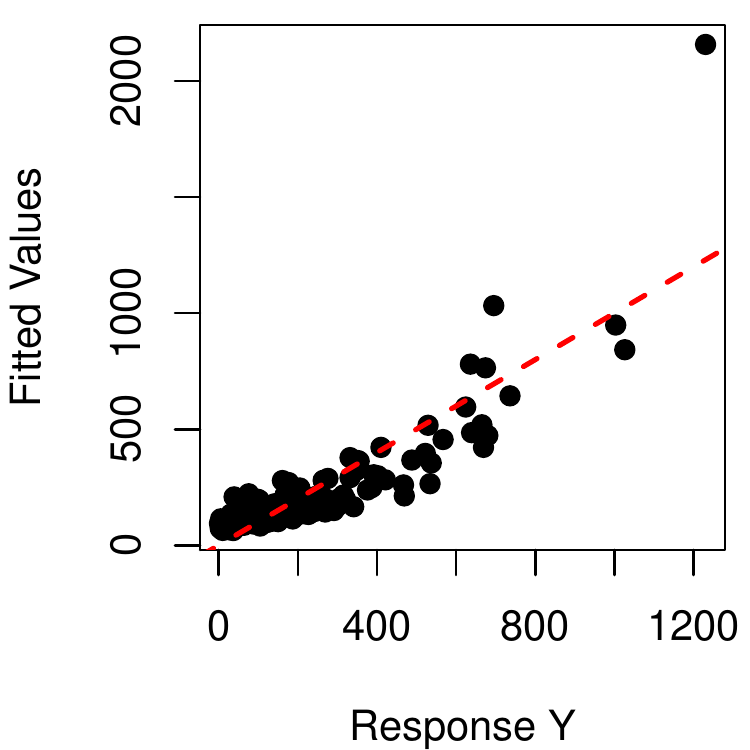} &
\includegraphics[width=0.3\textwidth]{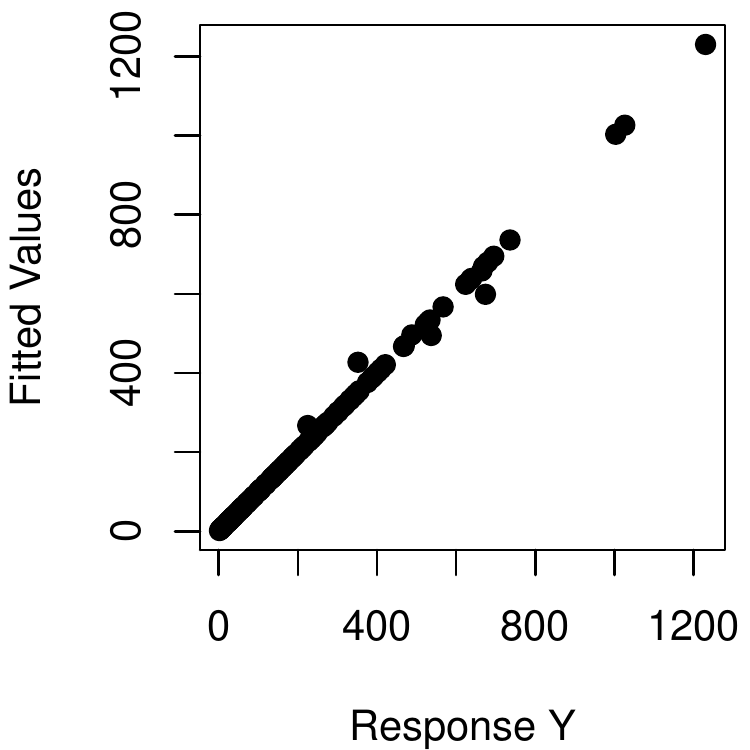} \\
(a) GLM & (b) MADE
\end{tabular}
\caption{Plots of the fittted response $\Yhat$ against the observed $Y$.}
\label{Fig:fishing}
\end{figure}

A Poisson generalized linear model fit gives a fit with an $R^2=\text{Cov}^2(Y, \Yhat)=0.75$. The fitted response is plotted against the observed response in Figure~\ref{Fig:fishing}(a). Ideally, we would expect these points to follow a 0-intercept unit-slope line (dashed line). The initial development of Poisson-MADE failed on this dataset. However, we addressed the issue by adding support for an offset, a fixed value to add to the regression function which is often used in count regression models. We fit a Poisson-MADE with $d=1$ and $h=169$. The estimated reduction is given in Table~\ref{Tab:fishing-est}. The result in Figure~\ref{Fig:fishing}(b) shows an excellent fit with $R^2=0.99$.

\begin{table}[t]
\centering
\small
\caption{Estimated reduction direction for fishing data.}
\label{Tab:fishing-est}
\tt
\begin{tabular}{rrrr}
&
density &
log(meandepth) &
log(sweptarea) \\
\hline
$\Bhat$ & 0.389 & -0.725 & -0.568 \\
\hline
\end{tabular}
\end{table}

\section{Discussion}
\label{sec:discuss}

We have introduced MADE as an extension to MAVE \citep{Xia02} for sufficient dimension reduction using local likelihood regression. While MAVE assumes a model consisting of a mean function with an additive error, MADE applies to exponential family outcomes such as Gaussian, Poisson, Binomial, etc. Initial simulations and data analyses have yielded encouraging results, but many issues can be raised and should be investigated in future work.

The present work does not offer analytical proof of the consistency of $\hat\alpha_j$, $\hat\gamma_j$, and $\Bhat$. A study of consistency and other statistical properties using the geometry of a Stiefel or Grassmann manifold would be of interest for MADE.

The performance of MADE depends on the bandwidth $h$. We have hand-picked $h$ to obtain results for the applications shown in this paper. We have also considered using a grid of possible $h$ values and selecting by cross-validation. This process becomes excessively computational because of the iterative estimation procedure, whose performance and reliability vary with the choice of $h$. We note that specific outcome types may admit faster and more efficient estimation procedures by foregoing the general exponential family framework; for example, closed-form solutions for $\hat\alpha_j$ and $\hat\gamma_j$ can be obtained in the Gaussian case.

We have devised and demonstrated permutation and bootstrap test procedures to determine the dimension of the reduction. We have also considered cross-validation, although its simulations were not reported. All three procedures are data-driven and computationally intensive. A possible alternative is to use an information theoretic criteria, such as Akaike's information criteria or the Bayesian information criteria. There is a limited literature on information criteria in local regression \citep{NonakaKonishi2005}, and we are not aware of an approach suitable for a parameter constrained to be a subspace or basis matrix. It may therefore be of interest to study information criteria which apply to parameters on manifolds.

So far, results and applications of the methodology were initiated for relatively smaller dimension of the predictors. Scalability of the methodology to large numbers of observations and dimensions should be investigated and properly evaluated. Group-wise reduction may help reduce the computing time along with a faster algorithm.

\appendix

\section{Appendix: Exponential Families}
\label{sec:exp-fams}
The MADE objective function \eqref{eqn:made-glm-obj} is formulated for outcomes from an exponential family. The function $b(\cdot)$ and canonical link $g(\cdot)$ are provided below for some selected exponential family distributions. To identify the canonical link between the mean parameter $\mu$ and the regression function $\vartheta(x)$, take Binomial as an example. The expression $\vartheta = \log[\mu / (1-\mu)]$ is multiplied by the observation $y/m$ in the exponential term, therefore $g(\mu) \stackrel{\text{def}}{=} \log[\mu / (1-\mu)] \equiv \text{logit}(\mu)$ is the canonical link function.
\begin{itemize}
\item $Y \sim \text{Binomial}(m, \mu)$ with known $m$
\begin{align*}
&f(y) = \binom{m}{y} \mu^y (1-\mu)^{m-y}
= \exp\left\{ \frac{(y/m) \log \frac{\mu}{1-\mu} + \log(1-\mu)}{1/m} \right\} \binom{m}{y} \\
&\vartheta = \log \frac{\mu}{1-\mu}, \quad
b(\vartheta) = \log(e^\vartheta + 1), \quad
b'(\vartheta) = \frac{1}{1 + e^{-\vartheta}}
\end{align*}

\item $Y \sim \text{Poisson}(\mu)$
\begin{align*}
&f(y) = \frac{e^{-\mu} \mu^y}{y!}
= \exp\left\{ y \log(\mu) -\mu \right\} \frac{1}{y!} \\
&\vartheta = \log \mu, \quad
b(\vartheta) = e^\vartheta, \quad
b'(\vartheta) = e^\vartheta
\end{align*}

\item $Y \sim \text{Geometric}(\mu)$
\begin{align*}
&f(y) = \mu (1-\mu)^y
= \exp\{ y \log(1-\mu) + \log \mu \} \\
&\vartheta = \log(1-\mu), \quad
b(\vartheta) = -\log(1-e^\vartheta), \quad
b'(\vartheta) = (e^{-\vartheta} - 1)^{-1}
\end{align*}

\item $Y \sim \text{NegBin}(\mu, \kappa)$ with known $\kappa$
\begin{align*}
&f(y) = \frac{\Gamma(y + \kappa^{-1})}{\Gamma(y+1) \Gamma(\kappa^{-1})}
\left(\frac{\kappa \mu}{1 + \kappa \mu}\right)^{y}
\left(\frac{1}{1 + \kappa \mu}\right)^{\kappa^{-1}} \\
&\quad= \frac{\Gamma(y + \kappa^{-1})}{\Gamma(y+1) \Gamma(\kappa^{-1})}
\exp\left\{
y \log \frac{\kappa \mu}{1 + \kappa \mu} +
\kappa^{-1} \log \frac{1}{1 + \kappa \mu}
\right\} \\
&\vartheta = \log \frac{\kappa \mu}{1 + \kappa \mu}, \quad
b(\vartheta) = \kappa^{-1} \log \frac{1}{1 - e^{\vartheta}}, \quad
b'(\vartheta) = \kappa^{-1} (e^{-\vartheta} - 1)^{-1}
\end{align*}

\item $Y \sim \text{N}(\mu, \sigma^2)$ with known $\sigma^2$
\begin{align*}
&f(y) = \frac{1}{\sigma \sqrt{2 \pi}} \exp\left\{ -\frac{1}{2 \sigma^2} (y - \mu)^2 \right\}
= \frac{1}{\sigma \sqrt{2 \pi}} \exp\left\{ \frac{-y^2}{2 \sigma^2} + \frac{y \mu - \mu^2/2}{\sigma^2} \right\} \\
&\vartheta = \mu, \quad
b(\vartheta) = \mu^2/2, \quad
b'(\vartheta) = \mu
\end{align*}

\item $Y \sim \text{Exp}(\mu)$
\begin{align*}
&f(y) = \frac{1}{\mu} e^{-y / \mu}
= \exp\{ -y / \mu + \log(1/\mu) \} \\
&\vartheta = -1/\mu, \quad
b(\vartheta) = -\log(-\vartheta), \quad
b'(\vartheta) = -1/\vartheta
\end{align*}

\item $Y \sim \text{Gamma}(\kappa, \mu)$ with known $\kappa$
\begin{align*}
&f(y) = \frac{y^{\kappa-1} e^{-y (\kappa / \mu)}}{\Gamma(\kappa) (\mu / \kappa)^{\kappa}}
= \exp\left\{ \frac{-\frac{1}{\mu} y - \log \mu}{1/\kappa} \right\} \frac{\kappa^{-\kappa} y^{\kappa-1}}{\Gamma(\kappa)} \\
&\vartheta = -1/\mu, \quad
b(\vartheta) = \log(-1/\vartheta), \quad
b'(\vartheta) = -1/\vartheta
\end{align*}

\item $Y \sim \text{InvGaussian}(\mu, \kappa)$ with known $\kappa$
\begin{align*}
&f(y) = \left( \frac{\kappa}{2 \pi y^3} \right)^{1/2} \exp\left\{ -\frac{\kappa (y - \mu)^2}{2 \mu^2 y} \right\}
= \exp\left\{ \frac{-\frac{y}{2 \mu^2} + \frac{1}{\mu}}{1/\kappa}  - \frac{\kappa}{2y} \right\} \left( \frac{\kappa}{2 \pi y^3} \right)^{1/2} \\
&\vartheta = -(-2\vartheta)^{-1/2}, \quad
b(\vartheta) = (-2\vartheta)^{1/2}, \quad
b'(\vartheta) = -(-2\vartheta)^{-1/2}
\end{align*}

\end{itemize}

\section{Appendix: Optimization on Stiefel manifold}
\label{sec:manifold-opt}

The collection of semi-orthogonal $p \times d$ matrices forms what is known as the Stiefel manifold. 
Working directly on the manifold acknowledges the constraints of the problem in a natural way. Optimization algorithms on manifolds require that the manifold is endowed with a differentiable structure so that fundamental operations, such as computation of a gradient or stepping from a previous iterate to the next iterate, are meaningful.

In a seminal paper on Stiefel and Grassmann manifold optimization of real-valued functions, \cite{EdelmanAS98} propose Newton-type and conjugate gradient algorithms. The algorithms rely on geodesics, tangent spaces, and other manifold constructs which are developed in that paper. We briefly summarize the conjugate gradient algorithm for Stiefel optimization used in the MADE algorithm for estimation of $B$, with some additional detail.
\begin{enumerate}
	\item Given $B_0$ such that $B_0^T B_0 = I$, compute $G_0 = F_{B_0} - B_0 F_{B_0}^T B_0$ and set $H_0 = -G_0$.
	\item For $k=0,1, \ldots$
	\begin{enumerate}
		\item Set $A = \left( B_k^T H - H^T B_k \right) / 2$.
		\item Calculate the norm of the gradient on the tangent space to $B_k$, equal to $\tr(H^T H) - \frac{1}{2}\tr(A^T A)$. If this norm is less than the tolerance, stop.
		\item Calculate the QR decomposition of $\left( I - B_k B_k^T \right) H_k$.
		
		\item Minimize $F(B_k(t))$ over t, where $B_k(t) = B_k M(t) + Q N(t)$, with $M(t)$ and $N(t)$ obtained by using the matrix exponential:
		$\left( \begin{array}{c}
		M(t) \\
		N(t)
		\end{array} \right)
		= \exp \left\{ t
		\left( \begin{array}{cc}
		A & -R^T \\
		R & 0
		\end{array} \right)
		\right\}
		\left( \begin{array}{c}
		I_d \\
		0
		\end{array} \right).
		$
		If $F(B_k) < F(B_k(t_{\text{min}}))$, shrink the search window.
		
		\item Set $B_{k+1} = B_k(t_{\text{min}})$.
		\item Compute $G_{k+1} = F_{B_{k+1}} - B_{k+1} F_{B_{k+1}}^T B_{k+1}$.
		\item Parallel transport tangent vector $H_k$ to the point $B_{k+1}$:
		\begin{equation*}
		\tau H_k = H_k M(t_{\text{min}}) - Y_k R^T N(t_{\text{min}}).
		\end{equation*}
		
		\item Use the conjugate gradient method to compute the new search direction, $H_{k+1} = -G_{k+1} + \gamma_k \tau H_k$, where
		\begin{equation*}
		\gamma_k = \frac{\langle G_{k+1} - \tau G_k, G_{k+1} \rangle}{\langle G_k, G_k \rangle}
		\end{equation*}
		with $\langle \Delta_1, \Delta_2 \rangle = \tr \left( \Delta_1^T \left( I - \frac{1}{2}Y_{k+1} Y_{k+1}^T \right) \Delta_2 \right)$. We use $\tau G_k = G_k$, although we may also use $\tau G_k = 0$.
		
		\item Reset $H_{k+1} = -G_{k+1}$ if $k+1 \equiv 0$ mod $d(p-d)+d(d-1)/2$.
	\end{enumerate}
\end{enumerate}
Note that the gradient for a function $F(B)$ with respect to the canonical metric on the Stiefel manifold is defined as $\nabla F = F_B - B F_B^T B$, where $F_B$ is the $p \times d$ matrix of partial derivatives of $F(B)$ with respect to the elements of $B$, i.e., $\left( F_B \right)_{rs} = \partial F(B)/\partial B_{rs}$. For the MADE objective function, the expression for $F_B$ is given in section~\ref{sec:estimate-coefs}. We refer interested readers to \cite{EdelmanAS98} and also \cite{OMM} for more information about matrix manifolds and optimization algorithms on matrix manifolds.


\end{document}